\documentclass[conference]{IEEEtran}

\usepackage{cite}
\usepackage{amsmath,amssymb,amsfonts}
\usepackage{graphicx}
\usepackage{textcomp}
\usepackage{xcolor}
\usepackage{tabularx}
\usepackage{booktabs}
\usepackage{url}

\begin{document}

\title{Name2Pkg: Lightweight One-Class Android Malware Screening via Name-Package Correspondence Modeling}

\author{
\IEEEauthorblockN{Changyeop Sung}
\IEEEauthorblockA{
\textit{School of Cybersecurity}\\
\textit{Korea University}\\
Seoul, Republic of Korea\\
scy6500@korea.ac.kr
}
\and
\IEEEauthorblockN{Yeonjae Kang}
\IEEEauthorblockA{
\textit{School of Cybersecurity}\\
\textit{Korea University}\\
Seoul, Republic of Korea\\
kangyj1995@korea.ac.kr
}
\and
\IEEEauthorblockN{Jaeho Shin}
\IEEEauthorblockA{
\textit{IT Planning Department}\\
\textit{Hana Bank}\\
Seoul, Republic of Korea\\
jaehoshin@hanafn.com
}
\and
\IEEEauthorblockN{Huy Kang Kim}
\IEEEauthorblockA{
\textit{School of Cybersecurity}\\
\textit{Korea University}\\
Seoul, Republic of Korea\\
cenda@korea.ac.kr
}
}

\maketitle

\begin{abstract}

Deep learning-based malware detection has been widely adopted in security-critical services. Most detection methods rely on internal features extracted from APK files or runtime behavior. However, extracting these features is computationally expensive. This limits their use in large-scale, early-stage screening. Malicious apps may exhibit weak correspondence between their user-facing app names and package names, providing a low-cost screening signal. We present Name2Pkg, a lightweight one-class classification method. It leverages only the app name and the package name. We formulate malware screening as a sequence anomaly detection problem. A character-level sequence-to-sequence model estimates the conditional likelihood of a package name given the app name. The length-normalized negative log-likelihood serves as the anomaly score. We train the model and calibrate the threshold using only benign data. Using a dataset of 67,129 real-world applications, Name2Pkg achieves an area under the receiver operating characteristic curve (ROC-AUC) of 0.982 and malware recall of 0.885 at an achieved false-positive rate of 0.044 on held-out test data. It has a 3.57 MiB checkpoint and a CPU inference latency of 28.20 ms per sample. Name2Pkg provides an efficient and effective pre-filtering signal for large-scale security systems.

\end{abstract}

\begin{IEEEkeywords}
Android malware screening, One-class anomaly detection, Name-package correspondence
\end{IEEEkeywords}

\section{Introduction}
Android malware detection commonly relies on static inspection of artifacts within Android application packages (APKs), dynamic analysis, or hybrid approaches
\cite{arp2014drebin,tam2015copperdroid,zhang2025mpdroid}.
Although these methods can achieve strong detection performance, they often require extracting permissions and API-call features, analyzing bytecode or graphs, or collecting runtime behavior. These requirements make them less suitable for large-scale triage and early-stage screening, where a low-cost signal is needed before deeper analysis. Prior work has therefore explored lightweight models, compact representations, and reduced feature sets
\cite{garcia2018revealdroid,kadir2025pacdroid,ma2024lightweight}.

App-market metadata and identifier strings provide another source of low-cost security signals \cite{teufl2016malware,martin2018android,munoz2015android}. SeqDroid, for example, learns representations from package names and other metadata strings together with permissions and intent actions \cite{lee2019seqdroid}. Studies of fake apps and app squatting also show that app names and package names can be manipulated or diverge in identity-abuse scenarios \cite{hu2020mobile,tang2019large}. This suggests that the relationship between the two identifiers may itself provide a screening signal. However, existing methods generally combine identifiers with additional features, treat them as independent attributes, or compare apps against known references. The correspondence between the app name and package name within a single app remains underexplored as a malware-screening signal.

In this paper, we focus on two low-cost textual identifiers: the app name and the package name. A package name is a namespace-style identifier commonly exposed through Android APIs and app metadata \cite{android_configure_app_module}, whereas the app name is the primary user-facing label. Because both identifiers can be obtained from app-market records or lightweight manifest-level metadata without bytecode analysis or runtime execution, they are natural inputs for low-cost early-stage screening.

A key challenge is that package names alone can already provide a useful signal. Some malicious apps use unusual, random-looking, or weakly meaningful package strings, and such patterns may be detected even without considering the app name. However, package-name morphology alone does not fully capture the relationship between what an app claims to be and how it is identified. A package name may look plausible in isolation but still be weakly related to the corresponding app name. Therefore, an identifier-level detector should model not only whether a package name looks normal, but also whether it is plausible given the app name. This restricted input also fits a one-class setting, in which benign regularities are learned without requiring complete or stable malware labels \cite{deloach2016android,wang2015accurate,wang2020evaluation,deldar2022android}.

We propose \textbf{Name2Pkg}, a lightweight one-class screening method based on name-conditioned package scoring. Name2Pkg uses a character-level sequence-to-sequence model trained only on benign app-name and package-name pairs
\cite{cho2014learning,sutskever2014sequence,bahdanau2015neural}.
Given an app name, the model estimates the conditional likelihood of the corresponding package name. The average negative log-likelihood of the package-name sequence is used as the anomaly score, with higher values indicating greater deviation from benign name-package regularities. The decision threshold is selected exclusively from a separate benign calibration set, and malware samples are used only for final evaluation. Name2Pkg is intended as a pre-analysis filter that prioritizes suspicious apps for more expensive static, dynamic, or hybrid analysis.

We evaluate Name2Pkg on a dataset collected by a commercial bank in South Korea, consisting of 62,730 benign apps and 4,399 malware samples. It achieves an area under the receiver operating characteristic curve (ROC-AUC) of 0.982 and, at a target false-positive rate (FPR) of 0.05, an achieved FPR of 0.044 on the benign test set and malware recall of 0.885. A package-only control reaches a ROC-AUC of 0.930 and recall of 0.692, showing that package-name morphology is useful but does not account for Name2Pkg's full performance. In a shuffled-pair control, breaking valid benign name-package correspondence increases anomaly scores. Name2Pkg has a 3.57 MiB checkpoint size and 28.20 ms CPU latency per sample.

The main contributions of this paper are as follows:

\begin{itemize}
    \item We formulate within-app name-package correspondence as a lightweight anomaly signal for Android malware screening.

    \item We develop Name2Pkg, a benign-only character-level sequence model that estimates conditional package-name likelihood and calibrates thresholds using benign data only.

    \item We evaluate Name2Pkg against identifier-level baselines, isolate the contribution of name-package correspondence through package-only and shuffled-pair controls, and report checkpoint size and CPU inference latency.
\end{itemize}

The remainder of this paper is organized as follows. Section II reviews related work. Section III details the Name2Pkg methodology. Section IV presents the experimental setup, evaluation results, and ablation studies. Finally, Section V concludes the paper.

\begin{figure*}[t]
    \centering
    \includegraphics[width=\textwidth]{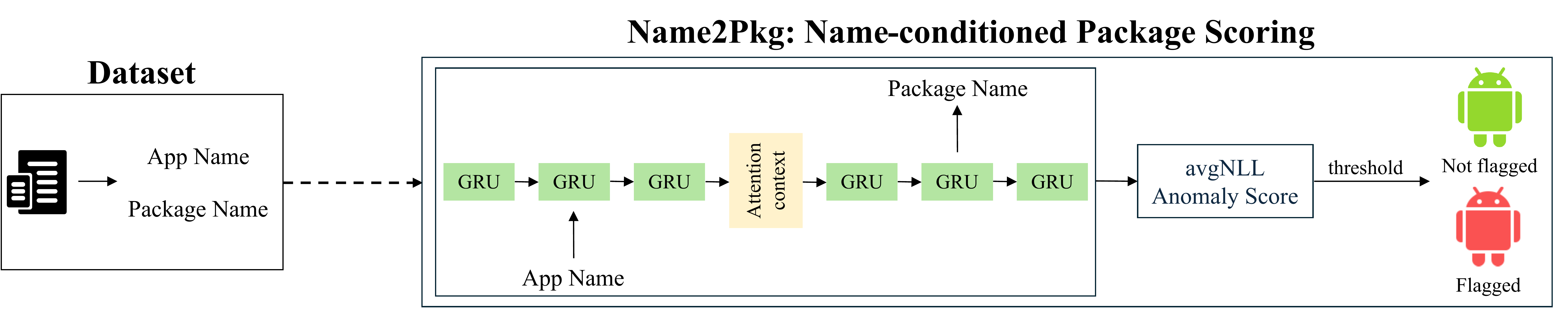}
    \caption{Overview of Name2Pkg. A character-level sequence-to-sequence model scores the observed package name conditioned on the app name. The length-normalized negative log-likelihood (avgNLL) serves as the anomaly score. Samples with scores at or above a benign-calibrated threshold are flagged for further analysis.}
    \label{fig:overview}
\end{figure*}

\section{Related Work}

\subsection{Static, Dynamic, and Hybrid Android Malware Detection}

Android malware detection has been extensively studied through static,
dynamic, and hybrid analyses of APK artifacts and runtime behavior
\cite{arp2014drebin,tam2015copperdroid,zhang2025mpdroid}.
Static methods inspect manifests, permissions, API calls, bytecode, or graph
representations without executing the app. DREBIN uses features extracted from
Android applications for static malware detection
\cite{arp2014drebin}, while GSEDroid represents apps using API-call graphs
enriched with permission and opcode-semantic information
\cite{gu2024gsedroid}.

Dynamic methods execute apps in controlled environments and observe behaviors
such as system calls, file and network operations, and sensitive API usage.
AppsPlayground and CopperDroid are representative systems for automated
runtime analysis and behavior reconstruction
\cite{rastogi2013appsplayground,tam2015copperdroid}.
Hybrid methods combine static and dynamic evidence. DeepAMD analyzes features
from both layers using a deep neural network
\cite{imtiaz2021deepamd}, whereas MPDroid integrates static and dynamic
representations within a multimodal framework
\cite{zhang2025mpdroid}.
Although these approaches provide rich structural and behavioral evidence,
they require APK-internal feature extraction, graph construction, controlled
execution, or a combination of these operations. Name2Pkg is intended as a
lower-cost pre-analysis signal that can be applied before such methods.

\subsection{Lightweight Android Malware Detection}

To reduce the cost of conventional analysis, prior work has explored compact
models and restricted feature sets. Krzyszton \textit{et al.}~developed an on-device
detector using features obtained through the Koodous platform
\cite{krzyszton2022lightweight}. RevealDroid uses a lightweight,
obfuscation-resilient representation for malware detection and family
identification \cite{garcia2018revealdroid}. Ma \textit{et al.}~proposed a lightweight
two-layer detection framework \cite{ma2024lightweight}, while PacDroid uses selected permissions, Intent Actions, and Intent Categories extracted from Android manifest files \cite{kadir2025pacdroid}.

These methods reduce computational cost by limiting model complexity or
narrowing the feature space. However, they generally continue to depend on
APK-derived information, including manifest entries, permissions, intents, API
usage, native-code indicators, or bytecode-level features. Name2Pkg targets a more restricted setting in which screening requires only the app name and package name. Its lightweight design therefore applies both to model size and to the amount of application information required at inference time.

\subsection{Metadata, Identifier-Level Signals, and App Identity Abuse}

App-market metadata can provide useful signals before detailed APK analysis.
Teufl \textit{et al.}~used descriptions, permissions, ratings, and developer information
for pre-installation malware detection
\cite{teufl2016malware}. Munoz \textit{et al.}~identified predictive Google Play
metadata, including developer, certificate, and intrinsic app features
\cite{munoz2015android}, and Martin \textit{et al.}~further demonstrated that market
metadata can support early-stage malware detection
\cite{martin2018android}. At the identifier-string level, SeqDroid shows that package names and certificate owner names can provide useful textual signals for obfuscated Android malware detection \cite{lee2019seqdroid}.

Research on camouflaged apps, fake apps, and app squatting further shows that
externally visible identifiers can be manipulated. Kywe \textit{et al.}~examined
camouflaged applications that imitate visible identity cues such as app names
and icons \cite{kywe2014camouflaged}. Hu \textit{et al.}~showed that both app names and
package names can be systematically manipulated in mobile app squatting
\cite{hu2020mobile}. Tang \textit{et al.}~found that fake apps frequently imitate
official app names but rarely reuse official package names
\cite{tang2019large}.

These studies establish the security relevance of metadata and identifier
strings, but they typically combine several fields, incorporate APK-derived
features, or compare an app with a known reference. Name2Pkg instead models the
relationship between the app name and package name within the same app. It
therefore requires neither an official reference app nor comparison against a
set of suspected clones.

\subsection{One-Class and Inconsistency-Based Android Malware Detection}

Benign-only, weak-label, and anomaly-based formulations have been studied to
reduce dependence on complete and reliable malware labels. Wang \textit{et al.}~used a
one-class support vector machine (SVM) trained on benign apps as part of a
cloud-based hybrid detection framework
\cite{wang2015accurate}. DeLoach \textit{et al.}~applied
positive–unlabeled learning to malware detection under weak ground truth
\cite{deloach2016android}. Wang and Zheng evaluated one-class feature-selection
and classification methods for zero-day detection using benign samples
\cite{wang2020evaluation}, while MalGAE learns benign attributed
function-call graphs using a stacked graph autoencoder
\cite{deldar2022android}.

A related line of work detects inconsistencies between an app's stated purpose
and its implementation. CHABADA groups apps according to description topics
and identifies API-usage outliers using one-class SVMs
\cite{gorla2014checking}. BERTDetect revisits this formulation using BERTopic
to model app descriptions and detect anomalous API usage
\cite{ranaweera2025bertdetect}.

These methods detect deviations between two information sources. However, they require full app descriptions and API-usage features. Name2Pkg restricts its input to two short identifier strings. It applies benign-only anomaly detection directly to within-app name-package correspondence.

\section{Methodology}
\label{sec:methodology}

Name2Pkg treats weak name-package correspondence as an identifier-level anomaly signal. The method learns regularities from benign
app-name and package-name pairs and then measures how unlikely a given package name is under benign correspondence patterns learned from benign data.

Name2Pkg uses a character-level sequence-to-sequence (Seq2Seq) model trained
only on benign pairs. Given a normalized app name, the model estimates the
conditional likelihood of the corresponding normalized package name. The average
negative log-likelihood of the package-name sequence is used as the anomaly
score. Higher scores indicate weaker name-package correspondence under benign
regularities. Figure~\ref{fig:overview} provides an overview of the screening process.

\subsection{Input Preprocessing}

The app name and package name are each normalized to reduce superficial formatting variation. For an app name $a$, spaces are inserted at CamelCase boundaries,
delimiters such as \texttt{.}, \texttt{\_}, and \texttt{-} are replaced with
spaces, consecutive spaces are collapsed, and the result is lowercased. For
example, \texttt{SampleTask\_App} is normalized to
\texttt{sample task app}.

For a package name $p$, underscores, hyphens, and spaces are replaced with
periods, repeated periods are collapsed, and the result is lowercased. For
example, \texttt{com.example.sample\_task} is normalized to
\texttt{com.\allowbreak example.\allowbreak sample.\allowbreak task}. The
normalized app name $\hat{a}$ is used as the source sequence, and the
normalized package name $\hat{p}$ is used as the target sequence.

\subsection{Name-conditioned Package Scoring}

Name2Pkg models name-package correspondence using a character-level
encoder-decoder architecture following prior sequence-to-sequence models
\cite{cho2014learning,sutskever2014sequence}. The encoder is a bidirectional
gated recurrent unit (GRU) over app-name characters, and the decoder is a
GRU that predicts package-name characters using additive
attention \cite{bahdanau2015neural}.

Given $\hat{a}$, the decoder predicts each package-name character conditioned on
the previous package-name characters and the encoded app-name representation.
Name2Pkg is trained by minimizing the average negative log-likelihood over benign training pairs. At inference time, the model scores the observed package-name sequence using its length-normalized negative log-likelihood (avgNLL):

\begin{equation}
S(\hat{a}, \hat{p})
=
-\frac{1}{|\hat{p}|}
\sum_{t=1}^{|\hat{p}|}
\log P_\theta(\hat{p}_t \mid \hat{p}_{<t}, \hat{a}).
\label{eq:name2pkg_score}
\end{equation}

Length normalization prevents longer package names from being penalized solely
because they contain more characters. Lower scores indicate plausible benign
name-package correspondence, whereas higher scores indicate greater deviation
from learned benign regularities.

\subsection{Benign-only Calibration and Decision Rule}

Name2Pkg follows a benign-only one-class calibration protocol. After training on
benign training samples, the anomaly scores of a separate benign calibration set
are used to select the decision threshold. Let
$\mathcal{S}_{\mathrm{cal}}^{B}$ denote the set of calibration scores computed
from benign calibration pairs. For a target false-positive rate $\gamma$, the threshold and decision rule are defined by

\begin{equation}
\begin{aligned}
\tau_\gamma
&=
Q_{1-\gamma}
\left(
\mathcal{S}_{\mathrm{cal}}^{B}
\right), \\
\hat{y}
&=
\mathbf{1}
\left\{
S(\hat{a}, \hat{p}) \geq \tau_\gamma
\right\}.
\end{aligned}
\label{eq:calibration_rule}
\end{equation}

Here, $Q_{1-\gamma}$ denotes the $(1-\gamma)$-quantile, while $\hat{y}=1$
denotes an anomalous sample and $\hat{y}=0$ denotes a non-anomalous sample. The threshold is determined entirely from benign
calibration scores; malware samples are not used for training, model selection,
or threshold calibration.

\section{Experiments}
\label{sec:experiments}

Under the benign-only one-class protocol, we evaluate Name2Pkg’s calibrated detection performance, comparisons against identifier-level baselines, computational efficiency, and the contribution of app-name conditioning.

\subsection{Dataset}

The evaluation uses an Android application dataset collected by a commercial
bank in South Korea. The benign samples were collected between March 2023 and
September 2023, whereas the malware samples were collected between October 2023
and June 2025. Each sample includes two textual identifiers used as input to Name2Pkg: the user-facing app name and the package name. Before filtering, the collected app names were written in a variety of scripts, including Latin, Hangul, CJK Unified Ideographs, Cyrillic, Katakana, and Hiragana. To ensure consistent character-level preprocessing across Name2Pkg and the identifier-level baselines, we retained apps whose names were primarily composed of ASCII Latin characters. Overall, preprocessing and filtering retained 62,730 of 115,388 benign samples and 4,399 of 12,992 malware samples, excluding 45.6\% and 66.1\%, respectively.

The benign data were split into training, calibration, and test sets at an 80/10/10 ratio. The calibration split was used for threshold selection, and the benign test split and all malware samples were held out for final evaluation. 
Table~\ref{tab:dataset_stats} summarizes the dataset split used for training, calibration, and evaluation.

\begin{table}[t]
\centering
\caption{SUMMARY OF DATASET. THE MODEL IS TRAINED AND CALIBRATED EXCLUSIVELY ON BENIGN SAMPLES.}
\label{tab:dataset_stats}
\setlength{\tabcolsep}{6pt}
\begin{tabular}{lc}
\toprule
\textbf{Dataset} & \textbf{Samples} \\
\midrule
Benign (train)       & 50,184 \\
Benign (calibration) & 6,273 \\
Benign (test)        & 6,273 \\
Malware (test)       & 4,399 \\
\midrule
Total                & 67,129 \\
\bottomrule
\end{tabular}
\end{table}

%
%
%

\subsection{Evaluation Metrics}

Name2Pkg is evaluated at benign-calibrated operating points with target false-positive rates $\gamma \in \{0.01, 0.05, 0.10\}$. For each method and each target FPR, the threshold is selected from benign calibration scores using
Eq.~\eqref{eq:calibration_rule}. Because the threshold is selected on
calibration data, the false-positive rate observed on the unseen benign test set
may differ from the target value. We therefore report the achieved
false-positive rate, denoted by aFPR.

We also report ROC-AUC, precision, recall, and F1-score. ROC-AUC is used as a
threshold-independent summary. Precision, recall, and F1-score are computed on
the combined evaluation set consisting of the benign test set and the malware
test set, whereas aFPR is computed only on the benign test set.

\begin{table*}[t]
\centering
\setlength{\tabcolsep}{3.5pt}
\caption{Performance comparison for each target FPR. aFPR denotes the achieved FPR on the benign
test set.}
\label{tab:main_results}
\begin{tabular}{lccccccccccccc}
\toprule
& & \multicolumn{4}{c}{\emph{Target FPR = 0.01}} 
& \multicolumn{4}{c}{\emph{Target FPR = 0.05}} 
& \multicolumn{4}{c}{\emph{Target FPR = 0.10}} \\
\cmidrule(lr){3-6}\cmidrule(lr){7-10}\cmidrule(lr){11-14}
\textbf{Method}
& \textbf{ROC-AUC}
& \textbf{aFPR} & \textbf{Prec.} & \textbf{Recall} & \textbf{F1}
& \textbf{aFPR} & \textbf{Prec.} & \textbf{Recall} & \textbf{F1}
& \textbf{aFPR} & \textbf{Prec.} & \textbf{Recall} & \textbf{F1} \\
\midrule
SBERT Top-1
& 0.903
& 0.011 & 0.820 & 0.070 & 0.128
& 0.055 & 0.828 & 0.378 & 0.519
& 0.103 & 0.805 & 0.605 & 0.690 \\

Package-only
& 0.930
& 0.010 & 0.977 & 0.607 & 0.749
& 0.051 & 0.906 & 0.692 & 0.784
& 0.098 & 0.846 & 0.766 & 0.804 \\

Conditional GRU-LM
& 0.948
& 0.008 & \textbf{0.981} & 0.622 & 0.761
& 0.049 & 0.912 & 0.727 & 0.809
& 0.098 & 0.856 & 0.826 & 0.840 \\

\textbf{Name2Pkg (ours)}
& \textbf{0.982}
& 0.010 & 0.979 & \textbf{0.656} & \textbf{0.786}
& 0.044 & \textbf{0.934} & \textbf{0.885} & \textbf{0.909}
& 0.097 & \textbf{0.875} & \textbf{0.962} & \textbf{0.916} \\
\bottomrule
\end{tabular}
\end{table*}

\subsection{Experimental Setup}
\label{sec:experimental_setup}

All experiments follow the benign-only protocol. All trainable models are trained only on the benign training split, thresholds are selected only from benign
calibration scores, and malware samples are used only for final evaluation.
Unless otherwise stated, the 80/10/10 benign split was generated using random seed 42.

\subsubsection{Shared Preprocessing and Vocabulary}

All methods use the normalization procedure described in
Section~\ref{sec:methodology}. App names and package names are truncated or
padded to a maximum length of 64 characters. The character vocabulary is built
only from benign training app names and package names to avoid leakage from
calibration, benign test, or malware test samples.

Name2Pkg uses \texttt{<PAD>}, \texttt{<UNK>}, \texttt{<BOS>}, and
\texttt{<EOS>} as special tokens. Package-only additionally uses
\texttt{<NULL\_APP>} as a fixed source token. Conditional GRU-LM uses
\texttt{<APP>}, \texttt{</APP>}, \texttt{<PKG>}, and \texttt{</PKG>} to mark
the app-name and package-name regions in the concatenated character sequence.

\subsubsection{Name2Pkg}

Name2Pkg is implemented as a character-level encoder-decoder Seq2Seq model.
The encoder is a one-layer bidirectional GRU over app-name characters, and the
decoder is a one-layer GRU over package-name characters with additive attention
over encoder hidden states. The character embedding dimension is 96, the encoder
hidden dimension is 192 per direction, and the decoder hidden dimension is 192.
Dropout is set to 0.1.

The model is trained with Adam using a learning rate of $10^{-3}$, batch size
256, and 8 epochs. Early stopping is not used. During training, the decoder uses
teacher forcing with the shifted package sequence. The anomaly score is computed
using Eq.~\eqref{eq:name2pkg_score}.

\subsubsection{Identifier-Level Baselines}

SBERT Top-1 is an embedding-based baseline using Sentence-BERT
\cite{reimers-2019-sentence-bert}, designed to measure semantic similarity
between an app name and the most similar package-name segment. We use the pretrained
\texttt{all-MiniLM-L6-v2} checkpoint \cite{allminilm}, based on the compact
MiniLM architecture \cite{wang2020minilm}, without task-specific fine-tuning.
The normalized package name is split into dot-delimited segments, and common
namespace tokens such as \texttt{com}, \texttt{org}, \texttt{net}, \texttt{kr},
\texttt{jp}, \texttt{io}, \texttt{co}, \texttt{de}, and \texttt{air} are
removed. The anomaly score is one minus the maximum cosine similarity between
the app-name embedding and the embeddings of the remaining package segments. Thresholds
are selected from benign calibration scores using the same protocol as
Name2Pkg.

Package-only is a Seq2Seq control with a fixed null app-name input. It uses the same Seq2Seq
architecture and hyperparameters as Name2Pkg, but replaces the app-name input
with the fixed token \texttt{<NULL\_APP>} during both training and inference.
Its anomaly score is the same average negative log-likelihood used in
Eq.~\eqref{eq:name2pkg_score}, with the fixed null source input. This control
tests whether package-name morphology alone can explain the observed detection
performance.

Conditional GRU-LM is a conditional GRU language model used as a simpler recurrent baseline.
Each input is represented as a single character sequence: app-name characters followed by package-name characters, with explicit delimiter tokens marking the regions. The model is a one-layer character-level GRU language model
with embedding dimension 96, hidden dimension 192, and dropout 0.1. It is
trained with Adam using a learning rate of $10^{-3}$, batch size 256, and 8
epochs. The loss and anomaly score are computed only over package-name character
positions after the \texttt{<PKG>} delimiter. This baseline tests whether a single recurrent conditional language model suffices without the explicit encoder–decoder structure used by Name2Pkg.

\subsubsection{Efficiency Measurement}

We measure single-sample CPU inference latency with a batch size
of 1 on an Intel Core i7-9700K CPU using 8 threads. Latency is averaged over
five runs on the same 500 apps, with 32 warm-up samples processed before each
run. Timing includes input encoding and scoring, but excludes shared string
normalization, loading, and disk I/O. Both Seq2Seq models compute encoder
attention projections once per input and reuse them across decoding steps.

\subsection{Main Detection Results}
\label{sec:main_results}

Table~\ref{tab:main_results} compares Name2Pkg with identifier-level baselines
and the package-only control under benign-calibrated operating points. Name2Pkg
achieves the strongest overall performance, with an ROC-AUC of 0.982. At a target FPR of 0.05, it achieves an aFPR of 0.044 on the benign test set,
with malware recall of 0.885. At this operating point, the calibrated threshold
is 2.408, yielding 3,893 true positives, 506 false negatives, 273 false positives,
and 6,000 true negatives. Precision and F1-score on this evaluation set are
0.934 and 0.909, respectively.

The comparison with Package-only shows that package-name morphology is
informative but insufficient. The package-only control reaches an ROC-AUC of 0.930 and recall of 0.692 at the target FPR of 0.05, indicating that anomalous
package names alone provide a meaningful signal. However, Name2Pkg improves
recall from 0.692 to 0.885 at the same target FPR. This gap supports our central hypothesis that conditioning on the app name
adds discriminative information beyond standalone package plausibility.

The remaining baselines clarify how this correspondence should be modeled.
Conditional GRU-LM also uses app-name context and outperforms the package-only
control, but it remains below Name2Pkg across the calibrated operating points.
This suggests that an attention-based encoder–decoder formulation is better suited for modeling name-conditioned package regularities than a single
recurrent language model over a concatenated sequence. In contrast, SBERT Top-1
performs substantially worse, especially at low false-positive budgets. This
indicates that off-the-shelf semantic similarity between an app name and package
segments does not adequately capture character-level identifier correspondence,
where package names often contain abbreviations, namespaces, developer tokens,
or short fragments.

Overall, the results show a consistent pattern: package-name morphology provides
a useful baseline signal, app-name conditioning strengthens that signal, and the
attention-based encoder-decoder formulation gives the strongest calibrated
identifier-level screening performance.

\subsection{Efficiency Analysis}
\label{sec:efficiency}

Table~\ref{tab:efficiency} reports the computational profiles of the identifier-level methods under single-sample CPU inference. The benchmark
includes input encoding and anomaly-score computation, but excludes shared
string normalization, loading, and disk I/O.

Name2Pkg requires a 3.57~MiB checkpoint and 28.20~ms per sample on CPU. Its
footprint is nearly identical to the package-only control because both use
the same Seq2Seq architecture; the main difference is whether the source input
is the actual app name or a fixed null token. The small latency gap between the
two models indicates that app-name conditioning adds limited inference overhead
relative to the package-only control.

Conditional GRU-LM is substantially smaller and faster, requiring 0.74~MiB and
6.57~ms per sample. However, this efficiency comes with lower calibrated
detection performance: at the target FPR of 0.05, its recall is 0.727 compared
with 0.885 for Name2Pkg. Name2Pkg has higher single-sample latency than SBERT Top-1 despite having fewer
parameters. Sequential decoding of package-name characters contributes to this
latency. SBERT Top-1 has a much
larger pretrained model footprint and substantially weaker recall under the
same benign-calibrated protocol. These results indicate that the key trade-off
is not latency alone, but whether the model captures the intended
name-package correspondence signal.

Name2Pkg is not the smallest or fastest method, but it provides the
best calibrated detection performance while remaining compact enough for
pre-analysis triage. Its compact checkpoint and measured CPU latency support its intended role as a
pre-analysis signal before more expensive static, dynamic, or hybrid analysis.

\begin{table}[t]
\centering
\caption{Model size and single-sample CPU inference latency. Latency is averaged over five runs.
Latency includes input encoding and scoring, but excludes shared string
normalization, loading, and disk I/O.}
\label{tab:efficiency}
\setlength{\tabcolsep}{5pt}
\begin{tabular}{lrrr}
\toprule
Method & Parameters & Checkpoint size & CPU latency \\
 &  & (MiB) & (ms/sample) \\
\midrule
SBERT Top-1        & 22,713,216 & 86.66 & 9.63 \\
Package-only   &    935,445 &  3.58 & 27.95 \\
Conditional GRU-LM &    191,894 &  0.74 & 6.57 \\
\textbf{Name2Pkg (ours)}           &    934,676 &  3.57 & 28.20 \\
\bottomrule
\end{tabular}
\end{table}

\begin{figure*}[t]
    \centering
    \includegraphics[width=0.90\textwidth]{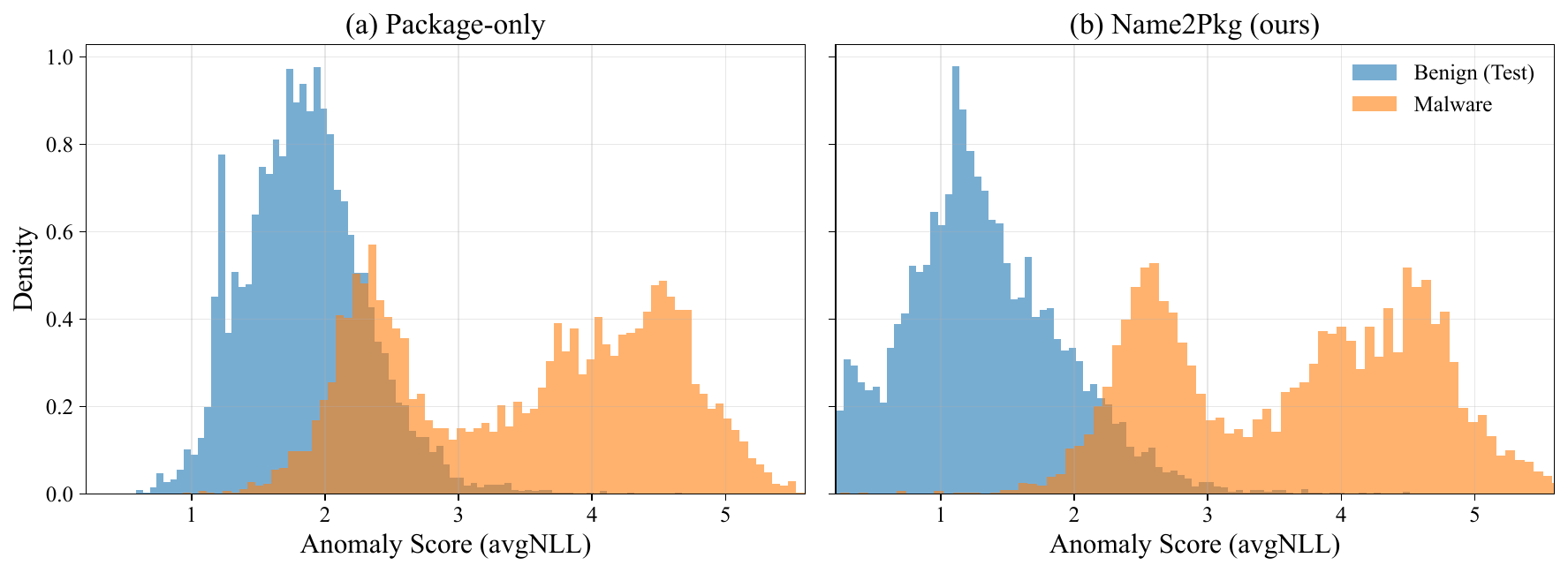}
    \caption{Anomaly-score distributions for the package-only control and
    Name2Pkg. Name2Pkg (right) reduces the overlap between benign and malware score distributions.}
    \label{fig:null_app_distribution}
\end{figure*}

\subsection{Ablation and Control Analysis}
\label{sec:ablation}

We further examine whether Name2Pkg exploits pairwise name-package
correspondence rather than relying only on package-name morphology. Because the
package-only control is already included in the main
comparison, this section focuses on score-distribution analysis and a name–package shuffle control.

\subsubsection{Distributional Effect of App-name Conditioning}

As described above, the package-only control isolates the effect of app-name conditioning by replacing the app-name input with a fixed null token.

Given the performance gap between Name2Pkg and the package-only control, we examine the score distributions to identify where app-name conditioning improves separation.

Figure~\ref{fig:null_app_distribution} compares the anomaly-score distributions
of the package-only control and Name2Pkg. Malware
scores are bimodal in both models. In the package-only control, the high-score
malware mode is already largely separated from benign samples, which is
consistent with package-name morphology providing a useful anomaly signal.
However, the low-score malware mode remains close to the benign distribution.
In the name-conditioned model, this low-score mode is more clearly separated
from benign samples.

The distributional comparison suggests that app-name conditioning adds a useful
signal where package-name morphology alone provides weaker separation. Together
with the package-only comparison, this supports the contribution of
name-conditioned package plausibility. We interpret this as distributional
evidence, not as evidence of fixed semantic categories among malware samples.

\subsubsection{Name-Package Shuffle Control}

We perform a name-package shuffle control on the benign test set. The Name2Pkg
model is trained on the original benign training pairs, and the thresholds are
selected from the original benign calibration set. We then randomly permute app
names within the benign test set while keeping package names fixed. The model is
not retrained or recalibrated.

This control preserves the marginal distributions of benign app names and
package names, but breaks their valid pairwise correspondence. If Name2Pkg
modeled only package-name morphology, this shuffle would have little effect on
the anomaly-score distribution. In contrast, a score increase on shuffled pairs
suggests that the model is sensitive to pairwise name-package correspondence.

Table~\ref{tab:shuffle_summary} summarizes the shuffle control. The mean anomaly
score increases from 1.326 for the original benign test pairs to 2.077 for the
name-shuffled pairs. The median score also increases from 1.254 to 2.077. This score increase suggests that breaking valid name–package correspondence makes otherwise benign package names less plausible to the name-conditioned model.

At the same time, the mean and median scores for the name-shuffled benign pairs
remain below the corresponding malware values of 3.632 and 3.785, respectively. This is expected because the shuffled samples still contain benign
package names, whereas malware samples may exhibit both unusual package-name morphology and weak name-package correspondence. Therefore, the shuffle control should be interpreted as a test of sensitivity to pairwise correspondence, not as a transformation
that makes benign samples equivalent to malware.

\begin{table}[t]
\centering
\caption{Name-package shuffle control on the benign test set. App names are
permuted while package names are fixed; the model is not retrained or
recalibrated. P95 and P99 denote the 95th and 99th percentiles, respectively.}
\label{tab:shuffle_summary}
\setlength{\tabcolsep}{5pt}
\begin{tabular}{lrrrrr}
\toprule
Set & Mean & Std. & Median & P95 & P99 \\
\midrule
Original benign & 1.326 & 0.593 & 1.254 & 2.360 & 2.909 \\
Shuffled benign & 2.077 & 0.473 & 2.077 & 2.851 & 3.336 \\
Malware & 3.632 & 0.978 & 3.785 & 5.049 & 5.447 \\
\bottomrule
\end{tabular}
\end{table}

Overall, the ablation and control results support a layered interpretation of
the proposed signal. Package names alone provide a meaningful baseline, but
conditioning on the app name improves detection by evaluating whether a package
name is plausible given the app name. The shuffle control further suggests that
Name2Pkg is sensitive to the relationship between the two identifiers rather
than merely learning the standalone plausibility of package strings.

\subsection{Discussion and Limitations}
\label{sec:discussion_limitations}

The results suggest that Name2Pkg captures two complementary identifier-level signals: standalone package-name regularity and name-conditioned package plausibility. The package-only control confirms that package-name morphology alone
carries useful information, while Name2Pkg's improvement over this control
shows that app-name conditioning adds further discriminative value. The comparison with Conditional GRU-LM also indicates that the modeling
formulation matters, suggesting that the attention-based encoder-decoder
architecture better captures the proposed correspondence signal. Finally, the score-distribution and shuffle
controls support the interpretation that Name2Pkg is sensitive to pairwise
name-package correspondence rather than only standalone package plausibility.

In deployment, Name2Pkg should be used as a triage filter: flagged apps can be
prioritized for deeper inspection, while unflagged apps receive lower priority.
Its decision relies only on identifier-level evidence, not on permissions, API
calls, bytecode, network behavior, native code, or runtime traces. This limited
input surface is central to its intended use: Name2Pkg provides a low-cost
screening signal before more expensive static, dynamic, or hybrid analysis is
applied. The evaluation set contains 41.2\% malware. Since precision and
F1-score depend on malware prevalence, these metrics may differ in deployment
environments with different class proportions.

As with other lightweight screening signals, Name2Pkg has a clear limitation under fully adaptive evasion. An attacker aware of the detection principle may choose an app name and package name that are mutually plausible
under benign naming conventions. For example, a malicious app may use a
banking-related app name together with a banking-related package name.
Such samples may receive low anomaly scores despite being malicious. This does not invalidate the intended use of Name2Pkg: its value lies in detecting
identifier-level irregularities at low cost and prioritizing suspicious apps
for deeper inspection, rather than serving as a standalone defense against
fully adaptive attackers.

False positives may also occur for legitimate apps with weakly aligned or
opaque identifiers. Some benign apps may use package names that reflect
internal project names, legacy namespaces, vendor identifiers, abbreviations,
or opaque organizational structures rather than the user-facing app name. Such
identifiers may be legitimate engineering or organizational artifacts, but they
can still appear anomalous to a name-conditioned model. In deployment, the decision threshold should therefore be selected according to the acceptable false-positive budget, as this setting controls how many legitimate but
weakly aligned apps are escalated for further analysis.

Because benign and malware samples were collected during different periods,
differences in collection periods may contribute to the observed separation
between the two classes. This study also restricts the input to app names written in Latin script.
This restriction provides a controlled setting for character-level preprocessing
and comparison with the English-centric SBERT baseline, while leaving the multilingual Android ecosystem for future work. App names written in non-Latin scripts, such as Hangul and CJK Unified Ideographs, may exhibit different naming patterns and different relationships with their package names. Even within
Latin-script app names, language-specific naming conventions may affect
name-package correspondence. Extending Name2Pkg to support multilingual app names, Unicode-aware preprocessing, and language-specific naming conventions is an important direction for future work.

\section{Conclusion}
\label{sec:conclusion}

This paper presented Name2Pkg, a lightweight anomaly detection method for Android malware screening that uses only the app name and package name. Name2Pkg employs a character-level sequence-to-sequence model and computes length-normalized negative log-likelihood scores.

The experimental results show that name-package correspondence is an effective
identifier-level screening signal. Name2Pkg achieves an ROC-AUC of 0.982 and, at a target FPR of 0.05, an aFPR of 0.044 on the benign test set with malware recall of 0.885. Compared with the package-only control, Name2Pkg improves the ROC-AUC from 0.930 to 0.982 and recall from 0.692 to 0.885 at the same target FPR, showing that app-name conditioning adds discriminative information beyond package-name morphology.

The ablation and control analyses further support this conclusion. The
distributional comparison shows that app-name conditioning improves separation
between benign apps and low-score malware samples, while the name-package
shuffle control shows that breaking valid benign pairings increases anomaly
scores. These findings suggest that Name2Pkg exploits pairwise name-package
correspondence rather than relying only on standalone package-name plausibility.

Name2Pkg also maintains a compact computational profile. It contains 934,676 parameters, has a checkpoint size of 3.57 MiB, and takes 28.20 ms per sample for CPU inference. Although it is not the smallest or fastest method
among the compared approaches, it provides the strongest calibrated detection
performance among the evaluated identifier-level methods while remaining
feasible for pre-analysis screening.

Overall, Name2Pkg demonstrates that lightweight identifier-level modeling can
provide a useful early signal for Android malware triage. Rather than replacing full static, dynamic, or hybrid malware analysis, Name2Pkg is designed to operate as a low-cost pre-analysis screening filter before such methods are applied. Apps whose name-package pairings are unlikely under benign naming regularities can then be prioritized for deeper inspection.

\bibliographystyle{IEEEtran}
\bibliography{references}

@inproceedings{arp2014drebin,
  author    = {Arp, Daniel and Spreitzenbarth, Michael and H{\"u}bner, Malte and Gascon, Hugo and Rieck, Konrad},
  title     = {{DREBIN}: Effective and Explainable Detection of {Android} Malware in Your Pocket},
  booktitle = {Proceedings of the Network and Distributed System Security Symposium ({NDSS})},
  year      = {2014},
  publisher = {Internet Society},
  doi       = {10.14722/ndss.2014.23247}
}

@inproceedings{rastogi2013appsplayground,
  author    = {Rastogi, Vaibhav and Chen, Yan and Enck, William},
  title     = {{AppsPlayground}: Automatic Security Analysis of Smartphone Applications},
  booktitle = {Proceedings of the Third {ACM} Conference on Data and Application Security and Privacy},
  pages     = {209--220},
  year      = {2013},
  publisher = {{ACM}},
  doi       = {10.1145/2435349.2435379}
}

@inproceedings{tam2015copperdroid,
  author    = {Tam, Kimberly and Khan, Salahuddin J. and Fattori, Aristide and Cavallaro, Lorenzo},
  title     = {{CopperDroid}: Automatic Reconstruction of {Android} Malware Behaviors},
  booktitle = {Proceedings of the Network and Distributed System Security Symposium ({NDSS})},
  pages     = {1--15},
  year      = {2015},
  publisher = {Internet Society},
  doi       = {10.14722/ndss.2015.23145}
}

@article{imtiaz2021deepamd,
  author    = {Imtiaz, Syed Ibrahim and ur Rehman, Saif and Javed, Abdul Rehman and Jalil, Zunera and Liu, Xuan and Alnumay, Waleed S.},
  title     = {{DeepAMD}: Detection and Identification of {Android} Malware Using High-Efficient Deep Artificial Neural Network},
  journal   = {Future Generation Computer Systems},
  volume    = {115},
  pages     = {844--856},
  year      = {2021},
  publisher = {Elsevier},
  doi       = {10.1016/j.future.2020.10.008}
}

@article{krzyszton2022lightweight,
  author    = {Krzyszto{\'n}, Mateusz and Bok, Bartosz and Lew, Marcin and Sikora, Andrzej},
  title     = {Lightweight On-Device Detection of {Android} Malware Based on the {Koodous} Platform and Machine Learning},
  journal   = {Sensors},
  volume    = {22},
  number    = {17},
  pages     = {6562},
  year      = {2022},
  publisher = {MDPI},
  doi       = {10.3390/s22176562}
}

@article{ma2024lightweight,
  author    = {Ma, Runze and Yin, Shangnan and Feng, Xia and Zhu, Huijuan and Sheng, Victor S.},
  title     = {A Lightweight Deep Learning-Based {Android} Malware Detection Framework},
  journal   = {Expert Systems with Applications},
  volume    = {255},
  pages     = {124633},
  year      = {2024},
  publisher = {Elsevier},
  doi       = {10.1016/j.eswa.2024.124633}
}

@article{kadir2025pacdroid,
  author    = {Kadir, Abdul and Peddoju, Sateesh Kumar},
  title     = {{PacDroid}: Lightweight {Android} Malware Detection Using Permissions and Intent Features},
  journal   = {Multimedia Tools and Applications},
  volume    = {84},
  number    = {27},
  pages     = {32351--32379},
  year      = {2025},
  publisher = {Springer},
  doi       = {10.1007/s11042-024-20455-w}
}

@inproceedings{wang2020evaluation,
  author    = {Wang, Yang and Zheng, Jun},
  title     = {An Evaluation of One-Class Feature Selection and Classification for Zero-Day {Android} Malware Detection},
  booktitle = {17th International Conference on Information Technology--New Generations ({ITNG} 2020)},
  series    = {Advances in Intelligent Systems and Computing},
  pages     = {105--111},
  year      = {2020},
  publisher = {Springer},
  doi       = {10.1007/978-3-030-43020-7_15}
}

@article{martin2018android,
  author    = {Mart{\'i}n, Ignacio and Hern{\'a}ndez, Jos{\'e} Alberto and Mu{\~n}oz, Alfonso and Guzm{\'a}n, Antonio},
  title     = {{Android} Malware Characterization Using Metadata and Machine Learning Techniques},
  journal   = {Security and Communication Networks},
  volume    = {2018},
  pages     = {5749481},
  year      = {2018},
  publisher = {Wiley},
  doi       = {10.1155/2018/5749481}
}

@incollection{lee2019seqdroid,
  author    = {Lee, William Younghoo and Saxe, Joshua and Harang, Richard},
  title     = {{SeqDroid}: Obfuscated {Android} Malware Detection Using Stacked Convolutional and Recurrent Neural Networks},
  booktitle = {Deep Learning Applications for Cyber Security},
  editor    = {Alazab, Mamoun and Tang, MingJian},
  pages     = {197--210},
  year      = {2019},
  publisher = {Springer},
  address   = {Cham},
  doi       = {10.1007/978-3-030-13057-2_9}
}

@inproceedings{reimers-2019-sentence-bert,
  author    = {Reimers, Nils and Gurevych, Iryna},
  title     = {{Sentence-BERT}: Sentence Embeddings Using Siamese {BERT}-Networks},
  booktitle = {Proceedings of the 2019 Conference on Empirical Methods in Natural Language Processing and the 9th International Joint Conference on Natural Language Processing ({EMNLP-IJCNLP})},
  pages     = {3982--3992},
  address   = {Hong Kong, China},
  publisher = {Association for Computational Linguistics},
  month     = nov,
  year      = {2019},
  doi       = {10.18653/v1/D19-1410},
}

@inproceedings{wang2020minilm,
  author    = {Wang, Wenhui and Wei, Furu and Dong, Li and Bao, Hangbo and Yang, Nan and Zhou, Ming},
  title     = {{MiniLM}: Deep Self-Attention Distillation for Task-Agnostic Compression of Pre-Trained Transformers},
  booktitle = {Advances in Neural Information Processing Systems},
  volume    = {33},
  pages     = {5776--5788},
  year      = {2020},
  publisher = {Curran Associates, Inc.},
  doi       = {https://doi.org/10.48550/arXiv.2002.10957}
}

@misc{android_configure_app_module,
  author       = {{Google}},
  title        = {Configure the app module},
  year         = {2026},
  howpublished = {\url{https://developer.android.com/build/configure-app-module}},
  note         = {Last updated: 2026-02-26; accessed: 2026-04-20}
}

@incollection{kywe2014camouflaged,
  author    = {Kywe, Su Mon and Li, Yingjiu and Deng, Robert H. and Hong, Jason I.},
  title     = {Detecting Camouflaged Applications on Mobile Application Markets},
  booktitle = {Information Security and Cryptology -- {ICISC} 2014},
  series    = {Lecture Notes in Computer Science},
  volume    = {8949},
  pages     = {241--254},
  publisher = {Springer},
  address   = {Cham},
  year      = {2015},
  doi       = {10.1007/978-3-319-15943-0_15}
}

@inproceedings{hu2020mobile,
  author    = {Hu, Yangyu and Wang, Haoyu and He, Ren and Li, Li and Tyson, Gareth and Castro, Ignacio and Guo, Yao and Wu, Lei and Xu, Guoai},
  title     = {Mobile App Squatting},
  booktitle = {Proceedings of The Web Conference 2020},
  pages     = {1727--1738},
  publisher = {{ACM}},
  year      = {2020},
  doi       = {10.1145/3366423.3380243}
}

@article{garcia2018revealdroid,
  author    = {Garcia, Joshua and Hammad, Mahmoud and Malek, Sam},
  title     = {Lightweight, Obfuscation-Resilient Detection and Family Identification of {Android} Malware},
  journal   = {{ACM} Transactions on Software Engineering and Methodology},
  volume    = {26},
  number    = {3},
  pages     = {1--29},
  year      = {2018},
  publisher = {{ACM}},
  doi       = {10.1145/3162625}
}

@inproceedings{deloach2016android,
  author    = {DeLoach, Jordan and Caragea, Doina and Ou, Xinming},
  title     = {{Android} Malware Detection with Weak Ground Truth Data},
  booktitle = {2016 {IEEE} International Conference on Big Data ({Big Data})},
  pages     = {3457--3464},
  year      = {2016},
  publisher = {{IEEE}},
  doi       = {10.1109/BigData.2016.7841008}
}

@article{wang2015accurate,
  author    = {Wang, Xiaolei and Yang, Yuexiang and Zeng, Yingzhi},
  title     = {Accurate Mobile Malware Detection and Classification in the Cloud},
  journal   = {SpringerPlus},
  volume    = {4},
  pages     = {583},
  year      = {2015},
  publisher = {Springer},
  doi       = {10.1186/s40064-015-1356-1}
}

@article{deldar2022android,
  author  = {Deldar, Fatemeh and Abadi, Mahdi and Ebrahimifard, Mohammad},
  title   = {{Android} Malware Detection Using One-Class Graph Neural Networks},
  journal = {The {ISC} International Journal of Information Security},
  volume  = {14},
  number  = {3},
  pages   = {51--60},
  year    = {2022},
  doi     = {10.22042/isecure.2022.14.3.6}
}

@article{teufl2016malware,
  author    = {Teufl, Peter and Ferk, Michaela and Fitzek, Andreas and Hein, Daniel and Kraxberger, Stefan and Orthacker, Clemens},
  title     = {Malware Detection by Applying Knowledge Discovery Processes to Application Metadata on the {Android Market} ({Google Play})},
  journal   = {Security and Communication Networks},
  volume    = {9},
  number    = {5},
  pages     = {389--419},
  year      = {2016},
  publisher = {Wiley},
  doi       = {10.1002/sec.675}
}

@inproceedings{munoz2015android,
  author    = {Mu{\~n}oz, Alfonso and Mart{\'i}n, Ignacio and Guzm{\'a}n, Antonio and Hern{\'a}ndez, Jos{\'e} Alberto},
  title     = {{Android} Malware Detection from {Google Play} Meta-Data: Selection of Important Features},
  booktitle = {2015 {IEEE} Conference on Communications and Network Security ({CNS})},
  pages     = {701--702},
  year      = {2015},
  publisher = {{IEEE}},
  doi       = {10.1109/CNS.2015.7346893}
}

@inproceedings{tang2019large,
  author    = {Tang, Chongbin and Chen, Sen and Fan, Lingling and Xu, Lihua and Liu, Yang and Tang, Zhushou and Dou, Liang},
  title     = {A Large-Scale Empirical Study on Industrial Fake Apps},
  booktitle = {2019 {IEEE}/{ACM} 41st International Conference on Software Engineering: Software Engineering in Practice ({ICSE-SEIP})},
  pages     = {183--192},
  year      = {2019},
  publisher = {{IEEE}},
  doi       = {10.1109/ICSE-SEIP.2019.00028}
}

@inproceedings{cho2014learning,
  author    = {Cho, Kyunghyun and van Merri{\"e}nboer, Bart and Gulcehre, Caglar and Bahdanau, Dzmitry and Bougares, Fethi and Schwenk, Holger and Bengio, Yoshua},
  title     = {Learning Phrase Representations Using {RNN} Encoder--Decoder for Statistical Machine Translation},
  booktitle = {Proceedings of the 2014 Conference on Empirical Methods in Natural Language Processing ({EMNLP})},
  pages     = {1724--1734},
  address   = {Doha, Qatar},
  publisher = {Association for Computational Linguistics},
  year      = {2014},
  doi       = {10.3115/v1/D14-1179},
}

@inproceedings{sutskever2014sequence,
  author    = {Sutskever, Ilya and Vinyals, Oriol and Le, Quoc V.},
  title     = {Sequence to Sequence Learning with Neural Networks},
  booktitle = {Advances in Neural Information Processing Systems},
  volume    = {27},
  pages     = {3104--3112},
  year      = {2014},
  publisher = {Curran Associates, Inc.},
  doi = {https://doi.org/10.48550/arXiv.1409.3215}
}

@inproceedings{bahdanau2015neural,
  author    = {Bahdanau, Dzmitry and Cho, Kyunghyun and Bengio, Yoshua},
  title     = {Neural Machine Translation by Jointly Learning to Align and Translate},
  booktitle = {3rd International Conference on Learning Representations ({ICLR} 2015)},
  year      = {2015},
  doi       = {https://doi.org/10.48550/arXiv.1409.0473}
}

@inproceedings{gorla2014checking,
  author    = {Gorla, Alessandra and Tavecchia, Ilaria and Gross, Florian and Zeller, Andreas},
  title     = {Checking App Behavior Against App Descriptions},
  booktitle = {Proceedings of the 36th International Conference on Software Engineering},
  pages     = {1025--1035},
  year      = {2014},
  publisher = {{ACM}},
  doi       = {10.1145/2568225.2568276}
}

@misc{allminilm,
  author       = {{Sentence-Transformers}},
  title        = {{all-MiniLM-L6-v2}},
  year         = {2025},
  howpublished = {\url{https://huggingface.co/sentence-transformers/all-MiniLM-L6-v2}},
  note         = {Model card; accessed: 2026-04-20}
}

@article{zhang2025mpdroid,
  author    = {Zhang, Sanfeng and Su, Heng and Liu, Hongxian and Yang, Wang},
  title     = {{MPDroid}: A Multimodal Pre-Training {Android} Malware Detection Method with Static and Dynamic Features},
  journal   = {Computers \& Security},
  volume    = {150},
  pages     = {104262},
  year      = {2025},
  publisher = {Elsevier},
  doi       = {10.1016/j.cose.2024.104262}
}

@inproceedings{ranaweera2025bertdetect,
  author    = {Ranaweera, Nishavi and Xu, Jiarui and Seneviratne, Suranga and Seneviratne, Aruna},
  title     = {{BERTDetect}: A Neural Topic Modelling Approach for {Android} Malware Detection},
  booktitle = {Companion Proceedings of the {ACM} on Web Conference 2025},
  pages     = {1802--1810},
  year      = {2025},
  publisher = {{ACM}},
  doi       = {10.1145/3701716.3717501}
}

@article{gu2024gsedroid,
  author    = {Gu, Jintao and Zhu, Hongliang and Han, Zewei and Li, Xiangyu and Zhao, Jianjin},
  title     = {{GSEDroid}: {GNN}-Based {Android} Malware Detection Framework Using Lightweight Semantic Embedding},
  journal   = {Computers \& Security},
  volume    = {140},
  pages     = {103807},
  year      = {2024},
  publisher = {Elsevier},
  doi       = {10.1016/j.cose.2024.103807}
}

\end{document}